\documentstyle[12pt,epsfig]{article}
\begin{document}
\rightline{September 2008}
\vskip 2cm
\centerline{
\large \bf
Early Universe cosmology in the light of the}
\vskip 0.3cm
\centerline{
\large \bf
mirror dark matter interpretation of the DAMA/Libra signal}
\vskip 1.7cm

\centerline{P. Ciarcelluti \footnote{paolo.ciarcelluti@ulg.ac.be}}
\vskip 0.5 cm
\centerline{
IFPA, D\'epartement AGO, Universit\'e de Li\`ege, 4000 Li\`ege, Belgium}

\vskip 1cm

\centerline{R. Foot \footnote{rfoot@unimelb.edu.au}}
\vskip 0.5 cm
\centerline{
School of Physics,
University of Melbourne, 3010 Australia.}

\vskip 1cm

\noindent
Mirror dark matter provides a simple framework for which to explain
the DAMA/Libra annual modulation signal consistently with the null
results
of the other direct detection experiments.
The simplest possibility involves ordinary matter interacting with
mirror dark matter via photon-mirror photon kinetic mixing of strength
$\epsilon \sim 10^{-9}$.
We confirm that photon-mirror photon mixing of this magnitude
is consistent with constraints from ordinary Big Bang nucleosynthesis 
as well as the more
stringent constraints from cosmic microwave background measurements and 
large scale structure considerations.

\newpage

A mirror sector of particles and forces can be well motivated from
fundamental considerations in particle physics, since its existence
allows for improper Lorentz symmetries, such as space-time parity and
time reversal, to be exact unbroken microscopic symmetries\cite{flv}.
The idea is to introduce a hidden (mirror) sector of particles and
forces, exactly duplicating the known particles and forces, except that 
in the mirror sector the roles of left and right chiral fields are
interchanged.
We shall denote the mirror particles with a prime $(')$.
In such a theory, the mirror protons and nuclei are naturally dark,
stable and massive, and provide an excellent candidate for dark matter
consistent with all observations and
experiments\cite{m0,m1,m2,m3,m4,m5,m6,Ciarcelluti-bbn,m7,m8,m8b,study,m9,m10,m11}.
For a review, see e.g. ref.\cite{review}.
Dark matter from a generic hidden sector is also possible, see e.g.
ref.\cite{feng} for a recent study.

It has been shown in ref.\cite{f08}, up-dating earlier
studies\cite{earlier}, that the mirror dark matter candidate is capable
of
explaining the positive dark matter signal obtained in the DAMA/Libra
experiment\cite{dama}, while also being consistent with the null results 
of the other direct detection experiments.
The simplest possibility sees the mirror particles coupling to the
ordinary particles via renormalizable photon-mirror photon kinetic
mixing\cite{he} (such mixing can also be induced radiatively if heavy
particles exist charged under both ordinary and mirror
$U(1)_{em}$\cite{holdom}):
\begin{eqnarray}
{\cal L}_{mix} = {\epsilon \over 2}F^{\mu \nu} F'_{\mu \nu}
\end{eqnarray}
where $F^{\mu \nu} = \partial^{\mu} A^{\nu} - \partial^{\nu}A^{\mu}$
and $F'^{\mu \nu} = \partial^{\mu} A'^{\nu} - \partial^{\nu}A'^{\mu}$.
This mixing enables mirror charged particles to couple to ordinary
photons
with charge $\epsilon qe$, where $q=-1$ for $e'$, $q=+1$ for $p'$ etc.
The mirror dark matter interpretation of the DAMA/Libra experiment
requires\cite{f08} $\epsilon \sim 10^{-9}$, which is consistent with
laboratory and astrophysical constraints\cite{footrev}.

The purpose of this note is to study the implications of such mixing for
the early Universe.
In particular, we will check that this kinetic mixing is consistent with
constraints from ordinary Big Bang nucleosynthesis (BBN) as well as more
stringent constraints from cosmic microwave background (CMB) and large
scale structure (LSS) considerations.


In the mirror dark matter scenario, it is assumed 
there is a temperature asymmetry ($T' < T$) between the ordinary and mirror
radiation sectors in the early Universe
due to some physics at early times (for specific models,
see e.g. \cite{kst}).
This is required in order to explain ordinary BBN, which suggests that
$T'/T \stackrel{<}{\sim} 0.6$.
In addition, several analyses\cite{m5,m6} based on numerical
simulations
of CMB and LSS suggest $T'/T \stackrel{<}{\sim} 0.3$.
However, if photon-mirror photon kinetic mixing exists, it can
potentially thermally populate the mirror sector.
For example, Carlson and Glashow\cite{CG} derived the approximate bound
of $\epsilon \stackrel{<}{\sim} 3 \times 10^{-8}$ from requiring that
the mirror sector does not come into thermal equilibrium with the
ordinary
sector, prior to BBN.
The inferred value of $\epsilon \sim 10^{-9}$ is consistent with 
this bound, so that
we
expect the kinetic mixing to populate the mirror sector, but with
$T'<T$. Assuming an effective initial condition $T' \ll T$, we can
estimate the
evolution of $T'/T$ in the early Universe as a function of $\epsilon$,
and thereby check the compatibility of the theory with the BBN and
CMB/LSS constraints on $T'/T$. 


Photon-mirror photon kinetic mixing can populate the mirror sector
in the early Universe via the process $e^+ e^- \to e'^+ e'^-$.
This leads to the generation of energy density in the mirror sector of:
\begin{eqnarray}
{\partial \rho' \over \partial t} =
n_{e^+} n_{e^-} \langle \sigma v_{M\o l} {\cal E} \rangle
\end{eqnarray}
where ${\cal E}$ is the energy transferred in the process,
$v_{m\o l}$ is the M\o ller velocity (see e.g. ref.\cite{gondolo}),
and $n_{e^-} \simeq n_{e^+} \simeq \frac{3\zeta(3)}{2\pi^2}T^3
$.

It is useful to consider the quantity: $\rho'/\rho$, in order
to cancel the time dependence due to the expansion of the Universe
[recall $\rho = \pi^2 g T^4/30$].
Using the time temperature relation:
\begin{eqnarray}
t = 0.3 g^{-1/2} {M_{Pl} \over T^2}
\end{eqnarray}
with $g = 10.75$ and $M_{Pl} \simeq 1.22 \times 10^{22}$ MeV, 
we find that:
\begin{eqnarray}
\frac{d(\rho'/\rho)}{dT} = \frac{-n_{e^-}n_{e^+} \langle \sigma
v_{M\o l} {\cal E} \rangle}
{\pi^2 g T^4/30} \ \frac{0.6 M_{Pl}}{\sqrt{g}T^3} \ .
\label{1}
\end{eqnarray}

Let us focus on $\langle \sigma v_{M\o l} {\cal E} \rangle$. This quantity
is:
\begin{equation}
\langle \sigma v_{M\o l} {\cal E} \rangle =
{\int \sigma v_{M\o l} (E_1 + E_2){1 \over 1 + e^{E_1/T}}
{1 \over 1 + e^{E_2/T}} d^3 p_1 d^3 p_2
\over
\int {1 \over 1 + e^{E_1/T}}
{1 \over 1 + e^{E_2/T}} d^3 p_1 d^3 p_2 }
\end{equation}
where we have neglected Pauli blocking effects.
If one makes the simplifying assumption of using Maxwellian
statistics
instead of Fermi-Dirac statistics then one can show (see appendix) that
in the massless electron limit:
\begin{equation}
\langle \sigma v_{M\o l} {\cal E} \rangle = {2\pi \alpha^2 \epsilon^2
\over 3T} \ ,
\label{ms}
\end{equation}
and
Eq.(\ref{1}) reduces to:
\begin{eqnarray}
{d(\rho'/\rho) \over dT} = {-A \over T^2}
\label{rho}
\end{eqnarray}
where
\begin{eqnarray}
A = {27\zeta(3)^2 \alpha^2 \epsilon^2 M_{Pl} \over \pi^5 g\sqrt{g} }\ .
\end{eqnarray}
Note that the $e'^{\pm}$ will
thermalize with $\gamma'$. However, 
because most of the $e'^{\pm}$ are produced in the low 
$T' \stackrel{<}{\sim}$ 5 MeV region, mirror weak interactions are too
weak to significantly populate the $\nu'_{e,\mu,\tau}$ [i.e.
one can easily verify a posteriori that the evolution of $T'/T$ for the
parameter space of interest is such that $G_F^2T'^5 \ll {\sqrt{g}T^2
\over 0.3 M_{Pl}}$].
Thus to a good approximation the radiation content of the mirror sector
consists of $e'^{\pm}, \gamma'$ leading to $g' = 11/2$ and hence
$\rho'/\rho = (g'/g)(T'^4/T^4)$,
with $g'/g \approx 22/43$.
 
Eq.(\ref{rho}) has the analytic solution:
\begin{eqnarray}
{T' \over T} = \left(\frac{g}{g'}A\right)^{1/4} \left[ {1 \over T} - {1 \over T_i}\right]^{1/4}
\label{ana}
\end{eqnarray}
where we have assumed the initial condition $T' = 0$ at $T = T_i$.

Let us now include the effects of the electron mass.
With non-zero electron mass, the evolution of $T'/T$ cannot be solved
analytically,
but Eq.(\ref{1}) can be solved numerically. Note that the number density
is:
\begin{eqnarray}
n_{e^-} = {1 \over \pi^2} \int^{\infty}_{m_e} { \sqrt{E^2 - m_e^2} E
\over  1 + exp(E/T) }  \ dE
\label{bla}
\end{eqnarray}
and, as we discuss in the appendix,
\begin{eqnarray}
\langle \sigma v_{M\o l} {\cal E} \rangle =
{1 \over 8m^4_e T^2 K_2^2 (m_e/T)} \int_{4m_e^2}^{\infty} ds \sigma (s -
4m_e^2)
\sqrt{s} \int_{\sqrt{s}}^{\infty} dE_+ e^{-E_+/T} E_+ \sqrt{{E_+^2 \over s}
- 1} 
\nonumber
\\
\
\label{bla2}
\end{eqnarray}
where the cross section is:
\begin{eqnarray}
\sigma = {4\pi \over 3} \alpha^2 \epsilon^2 {1 \over s^3} (s + 2m_e^2)^2
\ .
\end{eqnarray}
Numerically solving Eq.(\ref{1}) with the above inputs
(i.e. numerically solving the integrals Eq.(\ref{bla}) and
Eq.(\ref{bla2}) at each Temperature step), we find that\footnote{
For simplicity we have neglected the effect of heating 
of the photons via $e^{+} e^-$
annihilations. Note that the same effect occurs for the mirror photons
which are heated by the annihilations of $e'^+ e'^-$, so that $x_f$ is
approximately unchanged by this effect.}
\begin{eqnarray}
\epsilon  \simeq 8.5\times 10^{-10}  \left( {x_f \over 0.3}
\right)^2
\label{df3}
\end{eqnarray}
where $x_f$ is the final value ($T \to 0$) of $x = T'/T$.
In figure 1, we plot the evolution of $T'/T$,
for $\epsilon = 8.5 \times 10^{-10}.$

In deriving this result we have made several simplifying
approximations. The most significant of these are the following:
a) Using Maxwellian statistics instead of Fermi-Dirac statistics
to simplify the estimate of 
$\langle \sigma v_{M\o l} {\cal E} \rangle $. 
Using Fermi-Dirac statistics should decrease the interaction rate by around 8\%
as discussed in the appendix.
b) We have neglected Pauli blocking effects. Including Pauli blocking
effects will slightly reduce the interaction rate since some
of the $e'^{\pm}$ states are filled thereby reducing the available 
phase space. We estimate that the effect of the reduction of the 
interaction rate due to
Pauli blocking will be around
$\stackrel{<}{\sim}$ 10\%.
c) We have assumed that negligible $\nu'_{e,\mu,
\tau}$ are produced via mirror weak interactions from the $e'^{\pm}$.
Production of $\nu'_{e,\mu,\tau}$ will slightly decrease the
$T'/T$ ratio. The effect of this 
is equivalent to reducing
the interaction rate by around
$\stackrel{<}{\sim}$ 10\%. 
Taking these effects into account, we revise Eq.(\ref{df3}) to:
\begin{eqnarray}
\epsilon  = (1.0\pm 0.10)\times 10^{-9}  \left( {x_f \over 0.3}
\right)^2 \ .
\label{result}
\end{eqnarray}
Successful large scale structure studies\cite{m5,m6} suggest a rough
bound on $x_f$ of $x_f \stackrel{<}{\sim} 0.3$. Our result,
Eq.(\ref{result}), 
then suggests a rough bound on 
$\epsilon$ of $\epsilon \stackrel{<}{\sim} 10^{-9}$.

\vskip 0.6cm

\centerline{\epsfig{file=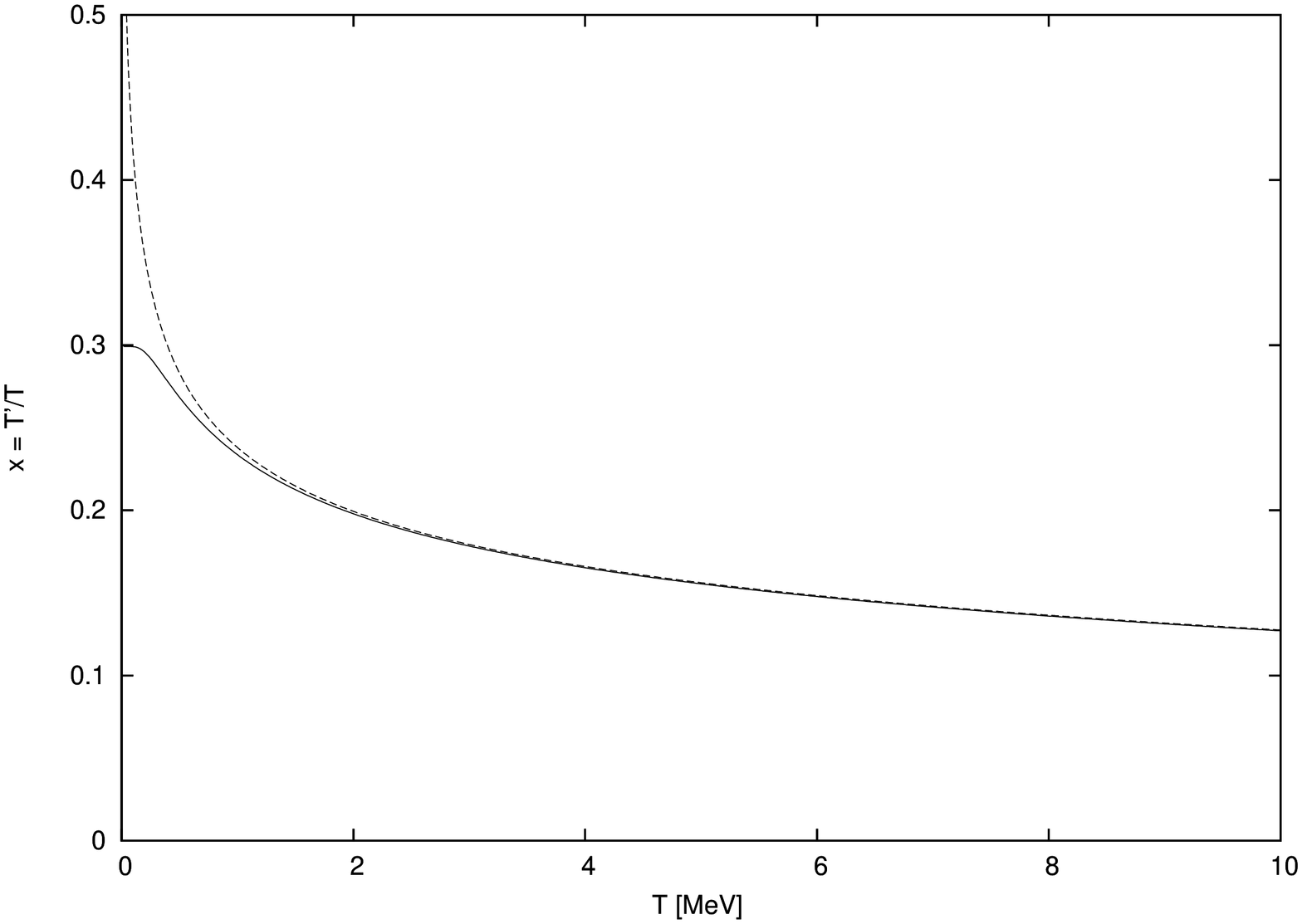,angle=270,width=12cm}}

\vskip 0.5cm
\noindent
Figure 1: Evolution of $x = T'/T$ for $\epsilon = 8.5 \times 10^{-10}$.
The solid line is the numerical solution including the effects of
the electron mass, while the dashed line is the analytic result [using
Eq.(\ref{ana}], which holds in the massless electron limit. As expected
the two solutions agree in the $T \stackrel{>}{\sim} 1 $ MeV region, where the
effects of the electron mass should be negligible.

\vskip 1cm

In conclusion, previous work has shown that the mirror dark matter
candidate can explain the DAMA/Libra annual modulation signal
consistently with the null results of the other direct detection
experiments provided that there exists photon-mirror
photon kinetic mixing of strength $\epsilon \sim 10^{-9}$.
Here we have examined the implications of this kinetic mixing for
early Universe cosmology, where we showed that it is consistent
with constraints from ordinary BBN and CMB/LSS data.

\vskip 1cm
\noindent
{\Large \bf Appendix}

\vskip 1cm

Here we shall examine the quantity 
$\langle \sigma v_{M\o l} {\cal E} \rangle$ and derive Eq.(\ref{ms})
and Eq.(\ref{bla2}) used in our analysis. 
Following ref.\cite{gondolo}, we have:
\begin{eqnarray}
\langle \sigma v_{M\o l} {\cal E} \rangle 
= { \int \sigma v_{M\o l} (E_1 +
E_2) e^{-E_1/T} e^{-E_2/T} d^3 p_1 d^3 p_2 \over
\int e^{-E_1/T} e^{-E_2/T} d^3 p_1 d^3 p_2 }
\end{eqnarray}
where $p_1$ and $p_2$ are the three-momenta and $E_1$ and $E_2$ the
energies of the colliding particles in the cosmic comoving frame.
Recall that ${\cal E} = E_1 + E_2$ is the energy transfer per collision.
As elaborated in ref.\cite{gondolo},
evaluation of these integrals can be facilitated by changing variables
to $E_{\pm} \equiv E_1 \pm E_2$ and $s = 2m_e^2 +
2E_1 E_2 - 2p_1 p_2 \cos\theta$. In terms of these variables the volume
element becomes
\begin{eqnarray}
d^3 p_1 d^3 p_2 = 2\pi^2 E_1 E_2 dE_+ dE_- ds
\end{eqnarray}
and
\begin{eqnarray}
\int \sigma v_{M\o l} {\cal E} e^{-E_1/T} e^{-E_2/T} d^3 p_1 d^3 p_2
=
2\pi^2 \int dE_+ E_+ \int dE_- \int ds \sigma v_{M\o l} E_1 E_2 e^{-E_+/T}
\nonumber \\
\ 
\end{eqnarray}
with integration region $|E_-| \le \sqrt{1 - \frac{4m^2_e}{s}}
\sqrt{E_+^2 -s}, \ E_+ \ge \sqrt{s}, \ s \ge 4m_e^2$.
Performing the $E_-$ integration, we have:
\begin{eqnarray}
\int \sigma v_{M\o l} {\cal E} e^{-E_1/T} e^{-E_2/T} d^3 p_1 d^3 p_2
= 4\pi^2 \int ds \sigma F \sqrt{1 - {4m_e^2 \over s}} \int dE_+
e^{-E_+/T} \sqrt{E_+^2 - s}\ E_+
\nonumber \\
\
\end{eqnarray}
where $\sigma F = \sigma v_{M\o l} E_1 E_2$ and $ F = {1 \over 2}
\sqrt{s(s-4m_e^2)}$.
Also, as discussed in ref.\cite{gondolo}
\begin{eqnarray}
\int e^{-E_1/T} e^{-E_2/T} d^3 p_1 d^3 p_2 &=& \left[ 4\pi m_e^2 T K_2
(m_e/T) \right]^2
\end{eqnarray}
where $K_2$ is the modified Bessel function of order 2.
Hence we see that
\begin{eqnarray}
\langle \sigma v_{M\o l} {\cal E} \rangle =
{1 \over 8m^4_e T^2 K_2^2 (m_e/T)} \int_{4m_e^2}^{\infty} ds \sigma (s -
4m_e^2)
\sqrt{s} \int_{\sqrt{s}}^{\infty} dE_+ e^{-E_+/T} E_+ \sqrt{{E_+^2 \over s}
- 1}  \ .
\nonumber
\\
\
\label{dfd}
\end{eqnarray}

In the $m_e \to 0$ limit, where $\sigma = {4\pi \alpha^2
\epsilon^2 \over 3s}$, and using the dimensionless
variable $z \equiv E_+/\sqrt{s}$, we find:
\begin{eqnarray}
\int \sigma v_{M\o l} {\cal E} e^{-E_1/T} e^{-E_2/T} d^3 p_1 d^3 p_2 &=&
{8\pi^3 \alpha^2 \epsilon^2 \over 3} 
\int^{\infty}_{0} ds s^{3/2} 
\int^{\infty}_{1} dz 
e^{-z\sqrt{s}/T}
z \sqrt{z^2 - 1}
\nonumber \\
& = & 128\alpha^2 \epsilon^2 \pi^3 T^5 I
\end{eqnarray}
where
\begin{eqnarray}
I \equiv \int^{\infty}_{1} {\sqrt{z^2 - 1} \over z^4} \  dz = {1 \over
3} \ . 
\end{eqnarray}
Also,
\begin{eqnarray}
\int e^{-E_1/T} e^{-E_2/T} d^3 p_1 d^3 p_2 &=& \left[ 4\pi m_e^2 T K_2
(m_e/T) \right]^2
\nonumber \\
&=& 64 \pi^2 T^6 \ {\rm in \ the \ m_e \to 0 \ limit.}
\end{eqnarray}
Thus we find:
\begin{equation}
\langle \sigma v_{M\o l} {\cal E} \rangle = {2\pi \alpha^2 \epsilon^2
\over 3T} \ .
\label{ms2}
\end{equation}

Our results for $\langle \sigma v_{M\o l} {\cal E} \rangle$,
Eq.(\ref{dfd}) [or Eq.(\ref{ms2}) for the $m_e \to 0$ limit], have assumed
Maxwellian distributions for the fermions to simplify the integrals.
In the $m_e \to 0$ limit, it is possible to evaluate the
integrals for the realistic case of Fermi-Dirac distributions.
In which case, one finds:
\begin{equation}
\langle \sigma v_{M\o l} {\cal E} \rangle =
{4\pi \alpha^2 \epsilon^2 \over 3T} {I_1 \over I_2}
\end{equation}
where
\begin{eqnarray}
I_1 &=& \int^{\infty}_0 dz \int^{\infty}_{\sqrt{z}} dx
\int^{\sqrt{x^2 - z}}_{0} dy
 \ {x \over 1 + e^{x+y}} {1 \over 1 + e^{x-y}}
\nonumber \\
I_2 &=&
\int^{\infty}_0 dz \int^{\infty}_{\sqrt{z}} dx
\int^{\sqrt{x^2 - z}}_{0} dy
 \ {x^2 - y^2 \over 1 + e^{x+y}} {1 \over 1 + e^{x-y}}
\end{eqnarray}
We find numerically that:
\begin{eqnarray}
I_1 \simeq 0.39, \ I_2 \simeq 0.84 \ \Rightarrow \ {I_1 \over I_2}
\simeq
0.46.
\end{eqnarray}
Thus, we see that the approximation of using Maxwellian statistics
overestimates 
$\langle \sigma v_{M\o l} {\cal E} \rangle $ by around 8\%.

\vskip 0.5cm
\noindent
{\bf Acknowledgements}
\vskip 0.2cm
\noindent
The work of R.F. was supported by the Australian Research Council.
P.C. acknowledges support from the Belgian fund for scientific
research (FNRS).

\end{document}